\documentclass{PoS}
\usepackage{lscape}
\usepackage{epsfig}
\usepackage{wrapfig}
\usepackage{subfigure}
\usepackage{amsmath}
\usepackage{hhline}
\usepackage{amssymb}
\usepackage{times}
\usepackage{cite}

\title{Inclusive diffraction at HERA}

\ShortTitle{Inclusive diffraction at HERA}

\author{\speaker{Laurent SCHOEFFEL} \\ %
        CEA Saclay, France \\
        E-mail: \email{laurent.schoeffel@cea.fr}}

\abstract{
A new experimental analysis of the 
diffractive process $ep \rightarrow eXY$,  where $Y$ denotes a proton or its low mass excitation with $M_Y<1.6$~GeV, has been performed with the H1 experiment at HERA \cite{Aaron:2012ad}.
The main results of this study are summarised in this document, together with the comparisons to other measurements and theoretical predictions. 
}

\FullConference{International Conference on the Structure and the Interactions of the Photon including the 20th International Workshop on Photon-Photon Collisions and the International Workshop on High Energy Photon Linear Colliders\\
                 20 - 24 May 2013\\
                 Paris, France}

\newcommand{\PO}{{I\!\!P}}
\newcommand{\RO}{{I\!\!R}}
\newcommand{\pom}{{I\!\!P}}

\def\gsim{\,\lower.25ex\hbox{$\scriptstyle\sim$}\kern-1.30ex%
\raise 0.55ex\hbox{$\scriptstyle >$}\,}
\def\lsim{\,\lower.25ex\hbox{$\scriptstyle\sim$}\kern-1.30ex%
\raise 0.55ex\hbox{$\scriptstyle <$}\,}

\newcommand{\xpom}{x_{\PO}}

\begin{document}

\section{Introduction}

At HERA a substantial fraction of up to $10\%$ of $ep$ interactions proceed via the diffractive scattering process initiated by a highly virtual photon~\cite{Aaron:2012ad,Aktas:2006hy,Aktas:2006hx,micha,Chekanov:2008cw,Chekanov:2008fh,FLDpaper}. 
In contrast to the standard deep inelastic scattering (DIS) process $e p\to e X$, the diffractive reaction $e p\to e X Y$ contains two distinct final state systems, where $X$ is a high-mass hadronic state and $Y$ is the elastically scattered proton or its low-mass excitation, emerging from the interaction with almost the full energy of the incident proton.

In the following, a new measurement of the diffractive neutral current DIS cross section is presented \cite{Aaron:2012ad}.
This is based upon H1 data for which there is an absence of hadronic activity in a large rapidity region extending close to the outgoing proton beam direction. 
The data were recorded with the H1 detector in the years 1999-2000 and 2004-2007, when HERA collided protons of $920$~GeV energy with $27.6$~GeV electrons and positrons.
The analysed data cover the low and medium $Q^2$ region from $3$ to $105$~GeV$^2$.
A combination with previous measurements obtained by H1, also using Large Rapidity Gap (LRG) events and based on low and medium $Q^2$ data from 1997 and high $Q^2$ data from 1999-2000~\cite{Aktas:2006hy}, is performed in order to provide a single set  of diffractive cross sections for $Q^2$ up to $1600$~GeV$^2$.

The study and interpretation of diffraction at HERA provides essential inputs for the understanding of quantum chromodynamics (QCD) at high parton densities.
The sensitivity of the diffractive cross section to the gluon density at low values of Bjorken $x$ can explain the high rate of diffractive events.
Diffractive reactions may therefore be well suited to search for saturation effects in the proton structure when $x$ reaches sufficiently small values~\cite{Marquet:2007nf}.

Several theoretical QCD approaches have been proposed to interpret the dynamics of diffractive DIS.
A general theoretical framework is provided by the QCD collinear factorisation theorem for semi-inclusive DIS cross sections such as that for $ep \rightarrow eXp$ \cite{collins,Trentadue:1993ka}.  
This implies that the concept of diffractive parton distribution functions (DPDFs) may be introduced, representing conditional proton parton probability distributions under the constraint of a leading final state proton with a
particular four-momentum.  
Empirically, an additional factorisation has been found to apply to good approximation, whereby
the variables which describe the proton vertex factorise from those describing the hard interaction (proton vertex factorisation)~\cite{is,is2}. 
The dependence of the DPDFs on the kinematic variables related to the proton vertex can be parametrised conveniently using Regge formalism, which amounts to a description of diffraction in terms of the exchange of a factorisable Pomeron ($\pom$)~\cite{pomeron} with universal parton densities. 
The experimental results described in this document are compared with QCD calculations based on DPDFs extracted from previous H1 data~\cite{Aktas:2006hy}.

\section{Diffractive DIS Kinematics Variables and Observables}

The kinematics of the inclusive DIS process can be described by the Lorentz invariants
\begin{equation}
x=\frac{-q^2}{2P \cdot q} , \hspace{1.5cm} y=\frac{P \cdot q}{P \cdot k} ,
\hspace{1.5cm} Q^2=-q^2 \ , 
\end{equation}
\noindent where $P$ and $k$ are the 4-momenta of the incident proton and electron\footnote{
In this paper the term ``electron'' is used generically to refer to both electrons and positrons.} respectively and $q$ is the 4-momentum of the exchanged virtual photon. 
The kinematics of the diffractive process can be described in addition by the invariant masses $M_X$ and $M_Y$ of the systems $X$ and $Y$, and
\begin{eqnarray}
t     &=&  (P-P_Y)^2 \ ,  \nonumber \\ 
 && \nonumber \\[-10pt]
\beta &=& \frac{-q^2}{2q \cdot (P-P_Y)} \ = \ \frac{Q^2}{Q^2+M_X^2-t} \ , \nonumber \\
 && \nonumber \\
x_{\PO}&=& \frac{q \cdot (P-P_Y)}{q \cdot P} 
\ = \ \frac{Q^2+M_X^2-t}{Q^2+W^2-m_P^2} \ = \ \frac{x}{\beta} \ ,
\label{eqkin}
\end{eqnarray}
\noindent where $P_Y$ is the 4-momentum of system $Y$, $W^2=(q+P)^2$ is the squared centre of mass energy of the virtual photon-proton system and  $m_P$ is the proton mass.
The variable $x_{\pom}$ is the fractional momentum loss of the incident proton.  
The quantity $\beta$ has the form of a Bjorken variable defined with respect to the momentum $P-P_Y$ lost by the initial proton.

In analogy to the inclusive DIS cross section, the inclusive diffractive cross section integrated over $t$ for $ep \to eXY$ in the one-photon exchange approximation can be written in terms of diffractive structure functions $F_2^{D(3)}$ and $F_L^{D(3)}$ as
\begin{equation}
\frac{{\rm d}^3\sigma^{ep \rightarrow eXY}}{
{\rm d}Q^2\, {\rm d}\beta\, {\rm d}x_{\pom}} = 
\frac{4\pi\alpha_{\mathrm{em}}^2}{\beta Q^4}
\biggl[ \Big(1-y+\frac{y^2}{2}\Big) F_2^{D(3)}(\beta,Q^2,x_{\pom}) 
  - \frac{y^2}{2} F_L^{D(3)}(\beta,Q^2,x_{\pom}) 
\biggr] ,
\label{sigma-2}
\end{equation}
\noindent where $\alpha_{\mathrm{em}}=1/137$. 
The structure function $F_L^{D(3)}$ corresponds to longitudinal polarisation of the virtual photon.
The reduced diffractive cross section is defined by
\begin{eqnarray}
\sigma_{r}^{D(3)}(Q^2,\beta,x_{\pom})&=& 
\frac{\beta Q^4}{4 \pi \alpha_{em}^2} \
\frac{1}{(1-y+\frac{y^2}{2})} \ \frac{{\rm d}^3 \sigma^{ep \rightarrow 
eXY}}{{\rm d}Q^2 \, {\rm d} \beta \, {\rm d} x_{\pom}} \\
&=&F_2^{D(3)}-\frac{y^2}{1+(1-y)^2}F_L^{D(3)} \ .
\label{sigred}
\end{eqnarray}

\section{Diffractive Cross Section Measurements and Combination}

Different event samples corresponding to different $Q^2$ ranges are analysed in this paper.  
For the interval $3 \le Q^2 \le 25 \rm\ GeV^2$, a `minimum bias' (MB) sample corresponding to an 
integrated luminosity of $3.5 \ {\rm pb^{-1}}$ is used, which was recorded during a special data taking period in 1999 with dedicated low $Q^2$ electron triggers. 
For  photon virtualities in the interval $10 \le Q^2 \le 105 \ {\rm GeV^2}$, data taken throughout the periods 1999-2000 and 2004-2007 are used, corresponding to a total integrated luminosity of $371 \ {\rm pb^{-1}}$. 
These cross section measurements are combined with previously published H1 LRG data~\cite{Aktas:2006hy}.

The 1999 MB, 1999-2000 and 2004-2007 data samples are used to measure the reduced diffractive cross section $\sigma_{r}^{D(3)}(Q^2,\beta,x_{\pom})$.
The bins in $Q^2$, $\beta$ and $x_{\pom}$ are chosen to have a width always larger than twice the experimental resolution. 
The cross section measurements are corrected to fixed values of $Q^2$, $\beta$ and $\xpom$ for each bin using predictions from the H1 $2006$ DPDF Fit B.
These corrections are of the order of $5\%$ in average.
The measurements are quoted at the Born level after correcting for QED radiative effects. 
Radiative corrections are calculated bin by bin using the HERACLES program~\cite{heracles} interfaced to RAPGAP.
They are smaller than $5\%$ for all measured data points. 
The results are corrected to the region $M_Y <1.6$~GeV, and $|t| \leq 1$~GeV$^2$.

Also, the new data sets of this analysis are combined with the previously published H1 measurements from the 1997 data~\cite{Aktas:2006hy} using the $\chi^2$ minimisation method developed for the combination of inclusive DIS cross sections~\cite{F2avGlazov,Aaron:2009bp,Aaron:2009wt}.
The combination is performed taking into account correlated systematic uncertainties.

\begin{figure}[!htbp]
\begin{center}
 \includegraphics[width=.7\textwidth]{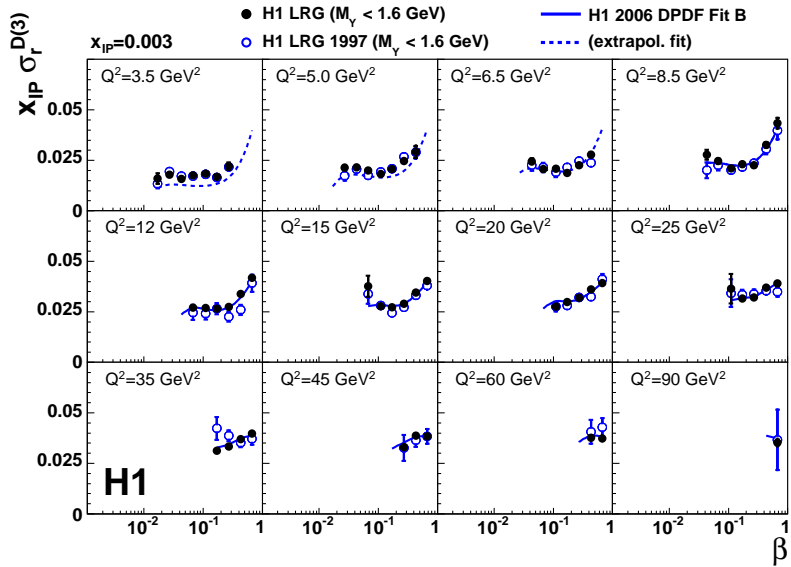}
\end{center}
  \caption{The $\beta$ dependence of the reduced diffractive cross section, multiplied by $\xpom$, at a fixed value of  $x_{\pom}= 0.003$, resulting from the combination of all data samples. 
Previously published H1 measurements~\cite{Aktas:2006hy} are also displayed as open points. 
The inner and outer error bars on the data points represent the statistical and total uncertainties, respectively. Overall normalisation uncertainties of $4\%$ and $6.2\%$ on the combined and previous data, respectively, are not shown.
Predictions from the H1 $2006$ DPDF Fit B~\cite{Aktas:2006hy} are represented by a curve in kinematic regions used to determine the DPDFs and by a dashed line in regions which were excluded from the fit.}
\label{fig:beta_dep3}
\end{figure}

The $\beta$ dependence of the combined reduced cross section measurements, multiplied by $x_{\pom}$, is shown in figure~\ref{fig:beta_dep3} for two fixed values of $x_{\pom}=0.003$ and $0.01$ are compared with the previously published cross section measurements~\cite{Aktas:2006hy} and with the prediction from the H1 $2006$ DPDF Fit B. 
A significant reduction of statistical errors is observed.
The new combined data have a total uncertainty between $4\%$ and $7\%$ whereas they were typically of the order of $7\%$ and $10\%$ in the previously published results.

A very good agreement with QCD calculations based on DPDFs extracted from previous H1 data~\cite{Aktas:2006hy} is obtained.

\begin{figure}[!htbp]
\begin{center}
 \includegraphics[width=.7\textwidth]{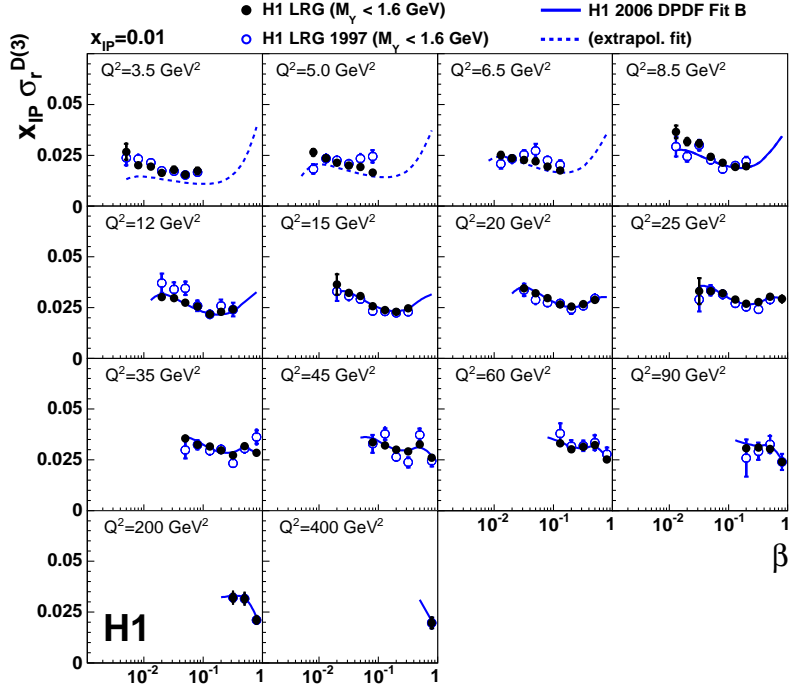}
\end{center}
  \caption{The $\beta$ dependence of the reduced diffractive cross section, multiplied by $\xpom$, at a fixed value of  $x_{\pom}= 0.01$, resulting from the combination of all data samples.}
\label{fig:beta_dep4}
\end{figure}

\section{Comparisons with other measurements}

The combined reduced cross section $\sigma_r^{D(3)}$ can be compared with other H1 measurements obtained by a direct measurement of the outgoing proton using the H1 Forward Proton Spectrometer (FPS)~\cite{micha}.
The cross section $ep \rightarrow eXY$ measured here with the LRG data includes proton dissociation to any system $Y$ with a mass in the range $M_Y < 1.6$~GeV, whereas in the cross section measured with the FPS the system $Y$ is defined to be a proton.
Since the LRG and FPS data sets are statistically independent to a large extent and the dominant sources of systematic errors are different, correlations between the uncertainties on the FPS and LRG data are neglected.
The ratio of the two measurements is then formed for each $(Q^2, \beta, x_{\pom})$ point for $\xpom=0.01$ and $\xpom = 0.03$, at which both LRG  and FPS data are available.
The global weighted average of the cross section ratio  LRG/FPS is
\begin{equation}
\frac{\sigma\left(M_Y < 1.6 \, {\rm GeV}\right)}{\sigma\left(Y=p \right)} = 1.203 \pm 0.019({\rm exp.}) \pm 0.087({\rm norm.}) \, ,
\label{eq:LRG_over_FPS}
\end{equation}
\noindent where the experimental uncertainty is a combination of statistical and uncorrelated systematic uncertainties on the measurements.

\begin{figure}[!htbp]
\begin{center}
 \includegraphics[width=.45\textwidth]{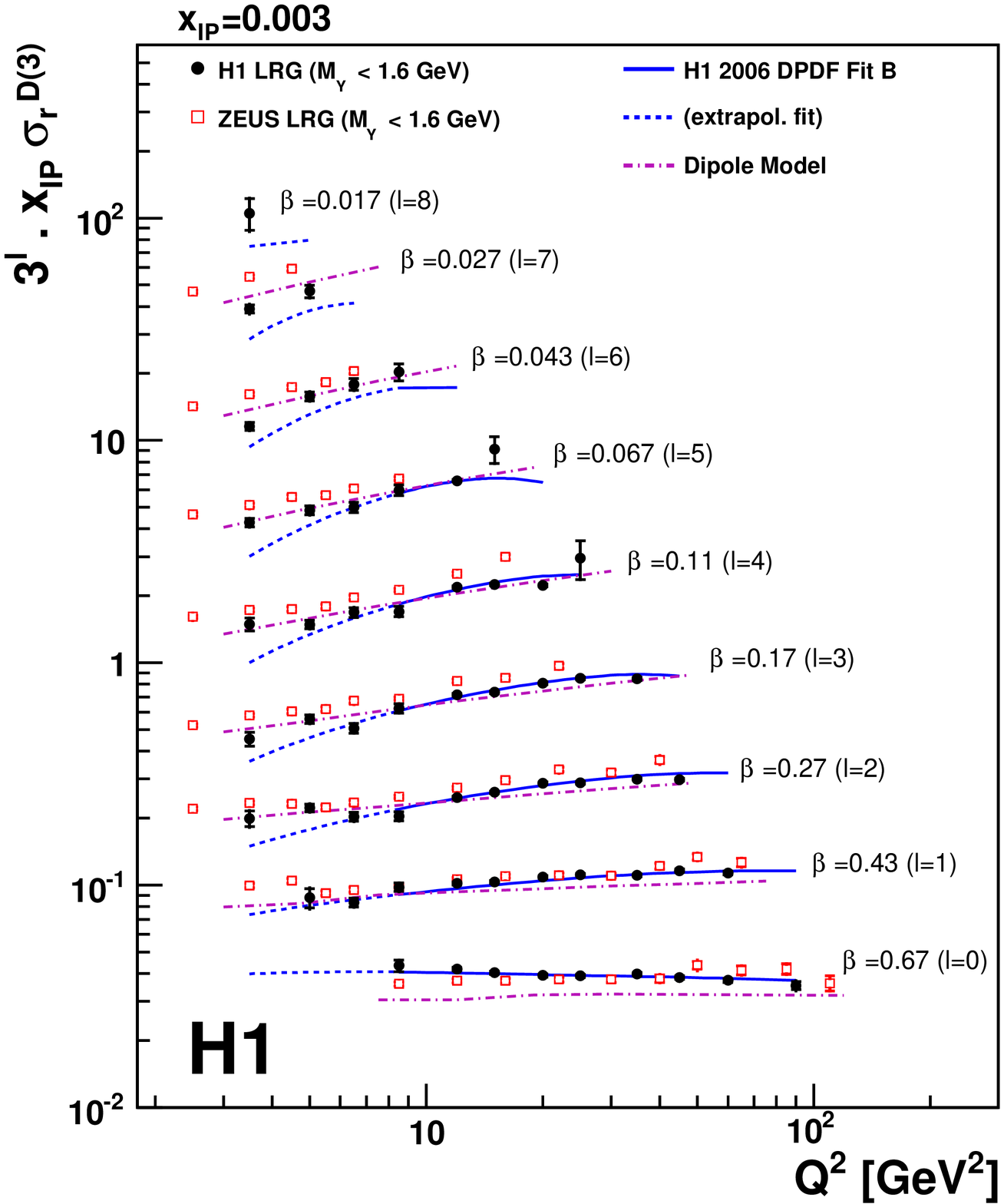}
 \includegraphics[width=.45\textwidth]{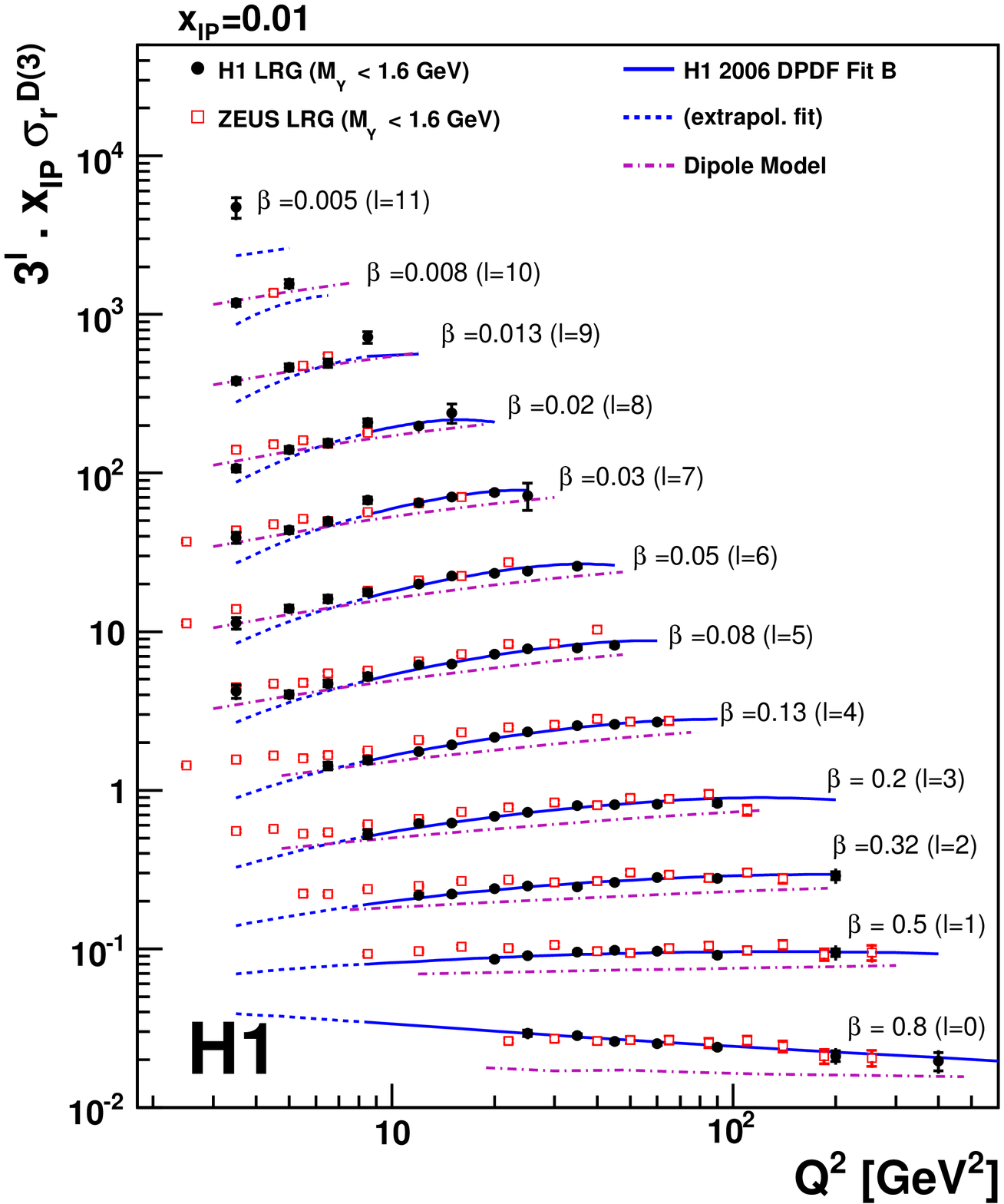}
\end{center}
  \caption{The $Q^2$ dependence of the
reduced diffractive cross section from combined H1 data,
multiplied by $\xpom$, at different fixed values of
$x_{\pom}$= $0.003$ and $0.01$. 
The present data are compared with the results of the ZEUS Collaboration~\cite{Chekanov:2008fh}, corrected to $M_Y < 1.6$~GeV (see text). 
The $8\%$ overall uncertainty on this correction for ZEUS data is not shown.
The overall normalisation uncertainties of $4\%$ and $2.25\%$ for the H1 and ZEUS data, respectively, are also not shown.}
\label{fig:ZEUS}
\end{figure}

The combined H1 LRG cross section are also compared with the most recent measurements by the ZEUS experiment using a similar LRG selection~\cite{Chekanov:2008fh}.
These ZEUS diffractive data have been determined for identical $\beta$ and $\xpom$ values, but at different $Q^2$ values to H1. 
In order to match the $M_Y < 1.6$~GeV range of the H1 data, a global factor of $0.91\pm 0.07$~\cite{Chekanov:2008fh} is applied to the ZEUS LRG data. 
The comparison for $M_Y < 1.6$~GeV between the H1 data and the rescaled ZEUS data is shown in figure~\ref{fig:ZEUS} for two values of $x_{\pom}$.
The ZEUS data tend to remain higher than those of H1 by $\sim 10\%$ on average.

This difference in normalisation is consistent with the $8\%$ uncertainty on the proton-dissociation correction factor of $0.91\pm 0.07$ applied to ZEUS data combined with the normalisation uncertainties of the two data sets of $4\%$ (H1) and $2.25\%$ (ZEUS).
This normalisation difference is also similar to that of $0.85$ $\pm$ $0.01$(stat.) $\pm$ $0.03$(sys.) $^{+0.09}_{-0.12}$(norm.) between the H1 FPS and the ZEUS LPS tagged-proton data sets~\cite{micha}. 
Deviations are observed between the $\beta$ dependencies of the two measurements at the highest and lowest $\beta$ values.
However a good agreement of the $Q^2$ dependence is observed throughout most of the phase space.

In addition, as discussed in the previous section, a good agreement with QCD calculations based on DPDFs extracted from previous H1 data~\cite{Aktas:2006hy} is obtained.

\section{Ratio to Inclusive DIS}

\begin{figure}[p]
  \center
  \includegraphics[width=.99\textwidth]{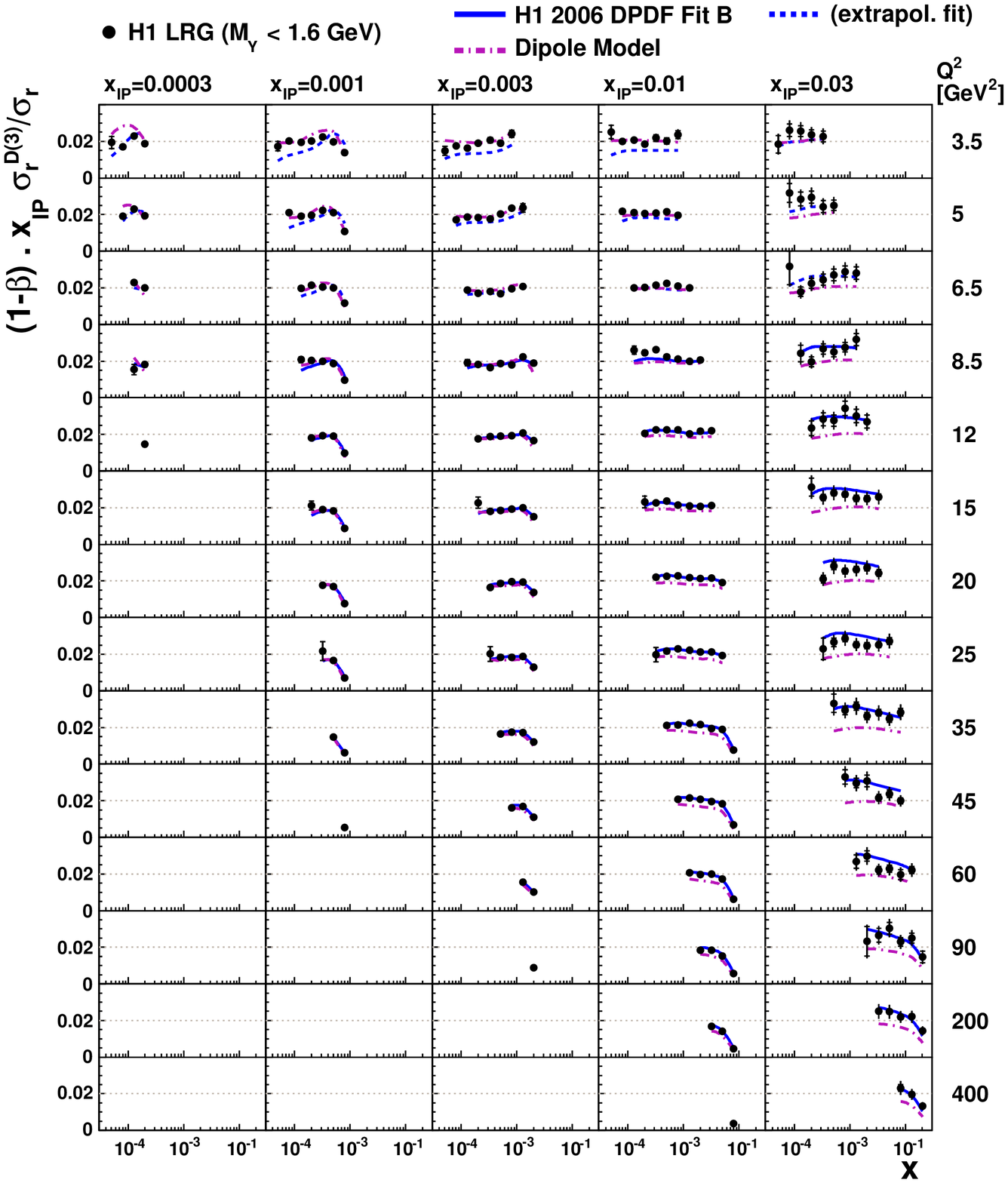}
  \caption{The ratio of the diffractive to the inclusive reduced cross section, multiplied by $(1-\beta)\xpom$. The inner and outer error bars on the data points represent the statistical and total uncertainties, respectively. The overall normalisation uncertainty of $4\%$ is not shown.}
  \label{fig:F2DoverF2}
\end{figure}

In analogy to hadronic scattering, the diffractive and the total cross sections can be related via the generalisation of the optical theorem to virtual photon scattering~\cite{optic_th}. 
Many models of low $x$ DIS~\cite{lowx1,lowx2,lowx3,lowx4,lowx5,lowx6} assume links between these quantities.
Comparing the $Q^2$ and $x$ dynamics of the diffractive with the inclusive cross section is therefore a powerful means of comparing the properties of the DPDFs with their inclusive counterparts and of testing models.
The evolution of the diffractive reduced cross section with $Q^2$ can be compared with that of the inclusive DIS reduced cross section $\sigma_r$ by forming the ratio 
\begin{equation}
\frac{\sigma^{D(3)}_r(x_\pom,x,Q^2)}{\sigma_r(x,Q^2)}\, . \, (1-\beta) \, x_\pom \, ,
\label{eq:diff_over_incl}
\end{equation} 
\noindent at fixed $Q^2$, $\beta = x/x_\pom$ and $x_\pom$. A parametrisation of $\sigma_r$ from~\cite{Aaron:2009kv} is used.
This quantity is equivalent to the ratio of diffractive to 
$\gamma^*p$~cross sections, 
\begin{equation}
\frac{M_X^2 \,\, \frac{\displaystyle {\rm d} \sigma_r^{D(3)}(M_X, W, Q^2)}{\displaystyle {\rm d} M_X}}{ \sigma_{incl.}^{\gamma^* p}(W,Q^2)} \, ,
\end{equation}
\noindent studied in~\cite{Chekanov:2008cw} as a function of $W$ and $Q^2$ in ranges of $M_X$.
Assuming proton vertex factorisation in the DPDF approach, this ratio is expected to be independent of $Q^2$ and  depends only weakly on $\beta$ and $x \simeq Q^2 / W^2$ for sufficiently large $M_X$. 
A remaining weak $x$ dependence of the ratio may arise due to deviations from unity of the intercept of the Pomeron trajectory, which are studied in the next section.
The ratio~(\ref{eq:diff_over_incl}) is shown in figure~\ref{fig:F2DoverF2} as a function of $x$ at fixed $x_\pom$ and $Q^2$ values.

The ratio of the diffractive to the inclusive cross section is found to be approximately constant with $x$ at fixed $Q^2$ and $x_\pom$ except towards larger $x$ values which correspond to large $\beta$ values. 
This indicates that the ratio of quark to gluon distributions is similar in the diffractive and inclusive process when considered at the same low $x$ value.
The ratio is also larger at high values of $x_\pom$, $\xpom = 0.03$, where the sub-leading exchange contribution of the diffractive cross section is not negligible, but it remains approximately constant with $x$.
The general behaviour of the ratio, and especially its decrease towards larger $x$, is reproduced by both the DPDF~\cite{Aktas:2006hy} predictions.

\section{Extraction of the Pomeron Trajectory}

The diffractive structure function $F_2^{D(3)}$ is obtained from the reduced cross section by correcting for the small $F_L^{D(3)}$ contribution using the predictions of the H1 $2006$ DPDF Fit B, which is in reasonable agreement with the recent direct measurement of $F_L^{D(3)}$~\cite{FLDpaper}. 
The diffractive structure function can be investigated in the framework of Regge phenomenology and is usually expressed as a sum of two factorised contributions corresponding to Pomeron and secondary Reggeon trajectories
\begin{eqnarray}
F_2^{D(3)}(Q^2,\beta,x_{\PO})=
f_{\PO / p} (x_{\PO}) \;  F_2^{\PO} (Q^2,\beta)
+n_{\RO}  \; f_{\RO / p} (x_{\PO}) \; F_2^{\RO} (Q^2,\beta) \ .
\label{reggeform}
\end{eqnarray}
\noindent In this parametrisation, $F_2^{\PO}$ can be interpreted as the Pomeron structure function  and $F_2^{\RO}$ as an effective Reggeon structure function.
The global normalisation of this last contribution is denoted $n_{\RO}$.
The Pomeron and Reggeon fluxes are assumed to follow a Regge behaviour with  
linear trajectories $\alpha_{\PO,\RO}(t)=\alpha_{\PO,\RO}(0)+\alpha^{'}_{\PO,\RO} t$, such that
\begin{equation}
f_{{\PO} / p,{\RO} / p} (x_{\PO})= \int^{t_{min}}_{t_{cut}} 
\frac{e^{B_{{\PO},{\RO}}t}}
{x_{\PO}^{2 \alpha_{{\PO},{\RO}}(t) -1}} {\rm d} t .
\label{flux}
\end{equation}
\noindent In this formula, $|t_{min}|$ is the minimum kinematically allowed value of $|t|$ and $t_{cut}=-1$ GeV$^2$ is the limit of the measurement.

In equation~(\ref{reggeform}), the values of $F_2^{\PO}$ are treated as  free parameters at each $\beta$ and $Q^2$ point, together with the Pomeron intercept $\alpha_{\PO}(0)$ and the normalisation $n_{\RO}$ of the sub-leading exchange.
The values of the other parameters are fixed in the fit.
The parameters $\alpha^{'}_{\PO}=0.04^{+0.08}_{-0.06}$~GeV$^{-2}$ and $B_{\PO}=5.7^{+0.8}_{-0.9}$ GeV$^{-2}$ are taken from the last H1 FPS publication~\cite{micha}.
The intercept of the sub-leading exchange $\alpha_{\RO}(0)=$~$0.5 \pm 0.1$ is considered.
The parameters $\alpha^{'}_{\RO}=0.30^{+0.6}_{-0.3}$~GeV$^{-2}$ and $B_{\RO}=1.6^{-1.6}_{+0.4}$~GeV$^{-2}$ are obtained from a parametrisation of previously published H1 FPS data~\cite{Aktas:2006hx}.
Since the sub-leading exchange is poorly constrained by the data, values of $F_2^{\RO} (Q^2,\beta)$ are taken from a parametrisation of the pion structure function~\cite{GRVpion}, with a single free normalisation $n_{\RO}$. 
Choosing a different parametrisation for the pion structure function~\cite{owens} does not affect the results significantly.
In previous publications~\cite{Aktas:2006hx,micha,Chekanov:2008fh}, it has already been shown that fits of this form provide a good description of the data. 
This supports the proton vertex factorisation hypothesis whereby the $x_{\pom}$ and $t$ dependences are decoupled from the $Q^2$ and $\beta$ dependences for each of the Pomeron and sub-leading contributions.
With the last measurements presented in this document, a new Regge analysis is performed.
Again, no significant $Q^2$ dependence of the Pomeron intercept is observed, which supports the proton vertex factorisation hypothesis.
The average value is found to be
\begin{eqnarray}
\alpha_{\PO}(0)= 1.113 \ \pm 0.002 \ \mathrm{(exp.)}  \ 
^{+ 0.029}_{- 0.015} \ \mathrm{(model)} \ , 
\label{alpom:answer}
\end{eqnarray}
\noindent where the first error is the full experimental uncertainty and the second error expresses the model 
dependent uncertainty arising dominantly from the variation of $\alpha_{\PO}^\prime$, which is strongly positively correlated with $\alpha_{\PO}(0)$. 
%

\section{Conclusions}

A measurement of the reduced inclusive diffractive cross section $\sigma_{r}^{D(3)}(Q^2,\beta,x_{\pom})$ for the process $ep \rightarrow eXY$ with $M_Y < 1.6 \ {\rm GeV}$ and $|t| < 1 \ {\rm GeV^2}$ as described in \cite{Aaron:2012ad} has been presented.
New results are obtained using high statistics data taken from 1999 to 2007 by the H1 detector at HERA.
These measurements are combined with previous H1 results obtained using the same  technique for the selection of large rapidity gap events.
The combined data span more than two orders of magnitude in $Q^2$ from $3.5 \ {\rm GeV^2}$ to $1600 \ {\rm GeV^2}$ 
and cover the range $0.0017 \leq \beta \leq 0.8$ for five fixed values of $\xpom$ in the range $0.0003 \leq \xpom \leq 0.03$. 
In the best measured region for $Q^2 \ge 12$ GeV$^2$, the statistical and systematic uncertainties are at the level of $1\%$ and $5\%$, respectively, with an additional overall normalisation uncertainty of $4\%$.
The combined H1 diffractive cross section measurements have been successfully compared with predictions from the DPDF approach.

\end{document}